\documentstyle[11pt,ysc,twoside,epsf]{article}
\def\edcomment#1{\iffalse\marginpar{\raggedright\sl#1\/}\else\relax\fi}
\reversemarginpar

\def\refSt{Starodubtseva et al., 2002}
\def\refHall{Hall, Riley, 1974}

\begin{document}

\title{Causes of Observed Long-Periodic Variations of the Polarization at Polar Regions of Jupiter}

\author{O. S. Shalygina,  V. V. Korokhin, E. V. Shalygin, G. P. Marchenko, Yu. I. Velikodsky, L. A. Akimov, O. M. Starodubtseva}
\affil{Astronomical Institute of Kharkov National University, Sumskaya Ul., 35, Kharkov, 61022, Ukraine. E-mail: dslpp@astron.kharkov.ua}

\begin{abstract}
Data of 23-years of Jupiter polarimetric observations (1981-2004)
have been reprocessed using new improved technique. The data from
other observers have been added to the analysis (1971-74). An
anticorrelation between asymmetry of polarization and insolation has
been found. The mechanism of influence of seasons' changing (through
temperature variations) on north-south asymmetry of polarization
formation has been proposed. Also a possibility of existence of
influence of solar cosmic rays flux on polarization value is noted.
\end{abstract}

\section{Seasonal variations of the north-south asymmetry of polarization}
Earlier on the basis of of 23-year (1981-2004) observational period
we have found the North-South asymmetry $P_N-P_S$
of linear polarization degree P and its seasonal variations (\refSt)
(parameter of asymmetry $P_N-P_S$ is a difference
between values of linear polarization degree on north and south at
the latitudes $\pm 60\deg$ at the central meridian). $P_N-P_S$
data are well organized if plotted in accordance with Jupiter's orbital
location and there is some relation between variations of polarization
and insolation (\refSt). In this work we are continuing our studying
the long-period variations of parameter $P_N-P_S$:
1) our new observations are used; 2) our old data (1981-1998) have
been reprocessed using new improved technique; 3) Hall and Riley data
(1971-1974) (ultraviolet, visual spectrum range) (\refHall) are involved
for analysis. New variant of P-asymmetry (on top) and intensity ratio
(below) dependences on Jupiter's orbital location are presented in
the fig.1.

\begin{figure}
\plotone{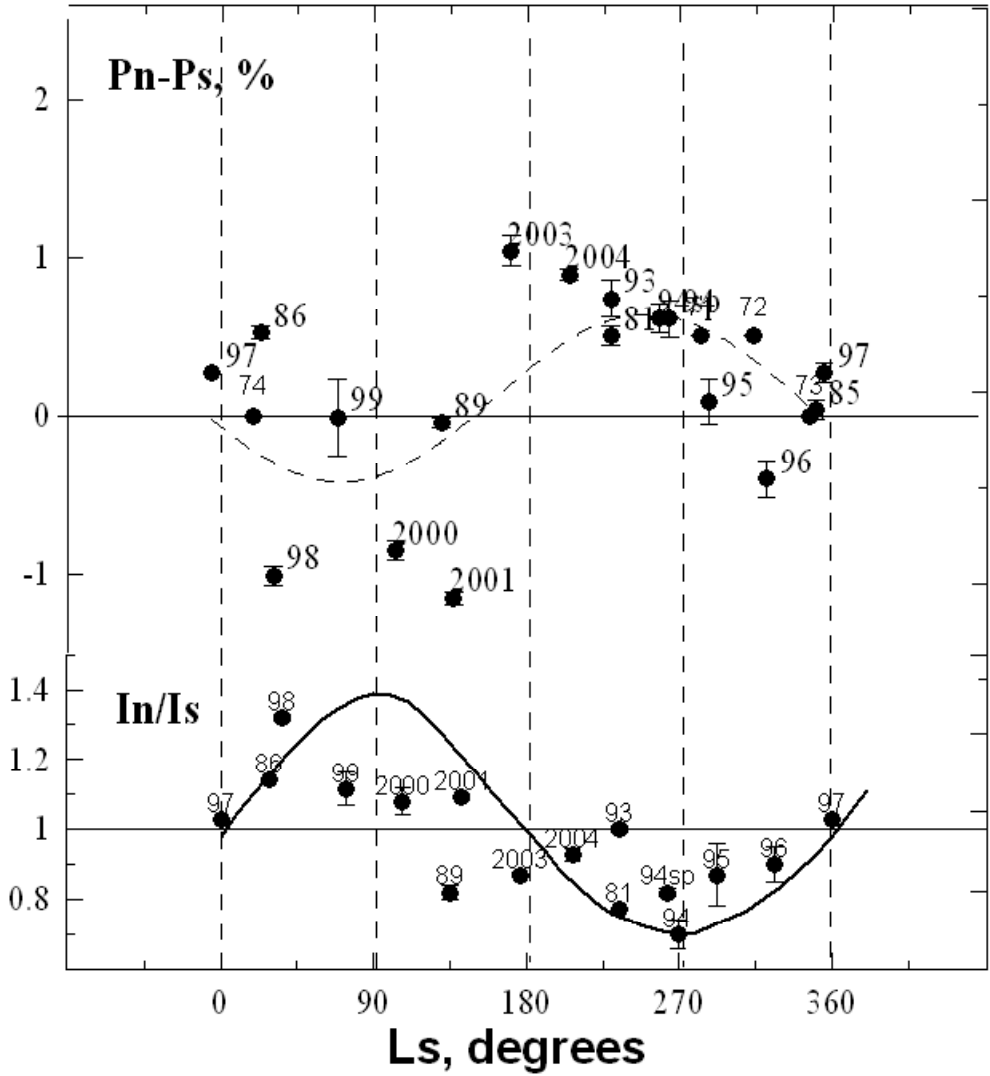}
\caption{Dependence of North-South asymmetry of polarization PN-PS (upper
plot) and intensity ratio $I_N/I_S$ (bottom plot) on planetocentrical orbital longitude of the Sun. Points are
the data obtained from our observations (in blue light). 1971-74 are the data of Hall and Riley  (\refHall) (ultraviolet, visual spectrum
range). Dotted curve (top plot) is approximating curve: $Y=0.53\sin(L_S-167.5\deg)+0.11$,
solid curve (bottom plot) is theoretically calculated asymmetry of insolation of
polar regions}
\end{figure}

\section{Investigation of nature of periodicity}
To investigate the nature of the dependence the approximations have
been made using different functions. The best approximation was
shown by one-periodic function (correlation coefficient is equal to
0.6. For comparison, correlation coefficients: for line
approximation is 0.23; for two-periodic function, which had been
used in previous publications (\refSt), is 0.31). Hall and Riley
data (\refHall) have a good agreement with our data. As one can see
(fig.1) there is an anticorrelation of long-term variations of
polarization and insolation.

\section{Causes of seasonal variations}
Thus, assumption about existence of the seasonal variations of
north-south asymmetry of polarization, proposed in our previous
papers, is correct. Moreover it is possible to assume that
variations of insolation are the principal cause of the seasonal
variations of polarization. How seasonal variations of insolation
may result in seasonal variations of polarization? We suppose that
by way of variations of temperature. Jupiter has a small axial tilt
(about 3\deg). However, the orbital eccentricity of 0.05 results in
20\% variation in the dilution factor $1/r^2$ values due to the
changing of the distance r from the Sun. Besides, the perihelion and
maximum of Jovian latitude of the Sun are almost coinciding in time.
These factors are causing the great seasonal fluctuations of the
incident solar radiation and causing the north-south asymmetry in
insolation and temperature. As a result, seasonal variations of
temperature are appear (mean temperature in Jupiter's atmosphere may
vary in the range $\pm 25$ K) (Beebe, Suggs \& Little 1986,
Caldwell, Cess \& Carlson, 1979). Which agent in Jupiter's
atmosphere may be sensitive to changes of temperature? Data of
polarimetric observations in visible, infrared and ultraviolet range
are sensitive to presence of stratospheric aerosols' haze in Jovian
atmosphere. This haze is located at the top pressure levels in the
range from a few mbar to a few tenths of mbar, with much more
abundance at high latitudes (latitudes greater than 40\deg-50\deg) .
Aerosols of this haze may be in unstable state and temperature (West
1988, Xanthakis et al., 1980) changing may influence on
generation/dissociation of particles. The anticorrelation of
polarization asymmetry and insolation may be caused by following
mechanism. Because of essential heating of thin stratospheric
aerosol layer (in Jovian summer) the substance of haze may leave
state of supersaturated vapor. Condensation become slower,
concentration of particles decreases and polarization also decreases
(as known, the rate of condensation decreases when temperature
increases). Thus, possible scenario of appearance of north-south
asymmetry of polarization is: seasonal variations of insolation
$\Rightarrow$ seasonal variations of temperature $\Rightarrow$
changes of activity of aerosol generation $\Rightarrow$ aerosol
concentration changes $\Rightarrow$ polarization changes
$\Rightarrow$ changes of north-south asymmetry of linear
polarization.

\section{Influence of solar activity on polarization changes}
 We investigated influence of solar wind, solar cosmic rays and X-rays on polarization
values. We have found that some correlation between $P_N-P_S$
and solar cosmic rays flux (protons, $E>10 MeV$) may exist (fig.2).
Let's pay attention to group of points, which marked on fig.2 (1998,
2000, 2001). These points greatly deviate just as on fig 1. Maybe,
extremely large flux of high-energy protons in this years (which has
been registered) had influenced on increasing of polarization values.

\begin{figure}
\plotone{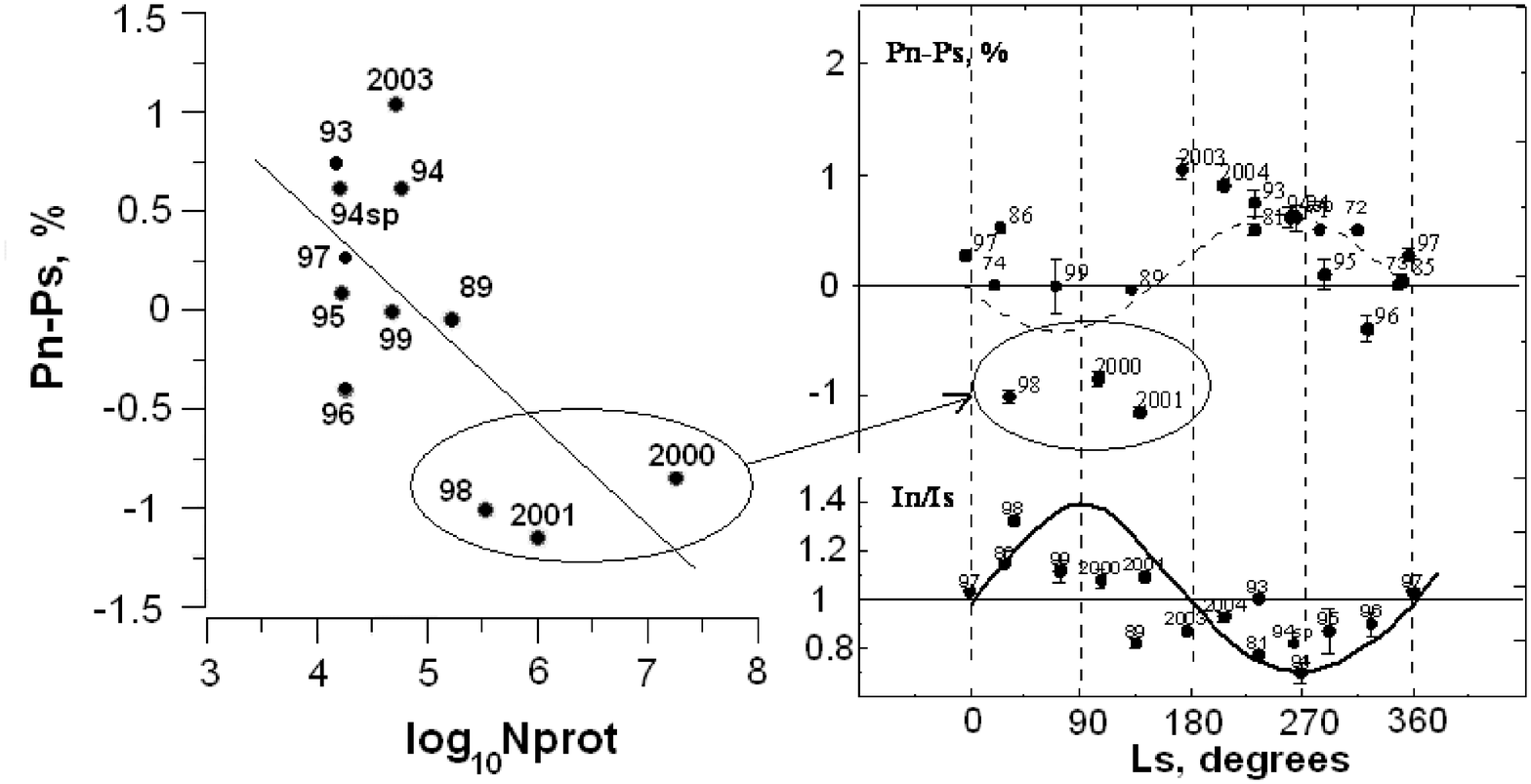}
\caption{Comparison of solar cosmic rays flux (amount of high energy
protons, GOES-10 data) with polarization asymmetry(left) and fig.1
(right)}
\end{figure}

Atmosphere temperature may be changed with increasing of amount of
high-energy protons in Jupiter's atmosphere, as it happens at the
Earth's stratosphere (Xanthakis J. et al. 1980). These changes may influence on aerosol
generation processes and be a cause of polarization asymmetry. Protons
of high energy may also be an additional centers of condensations
and thus may influence on aerosol concentrations and consequently
on polarization.

\section{Conclusions}
\begin{enumerate}
\item There is an anticorrelation between polarization asymmetry and insolation.
\item Seasonal variations of insolations (through variations of temperature)
is the principal cause of variations of north-south asymmetry of polarization.
\item Probably, there is some influence of solar cosmic rays flux (protons, $E>10 MeV$) on polarization value.
\end{enumerate}
You can find more details of this work at http://astron.kharkov.ua/dslpp/jup/.

\begin {references}
\reference Starodubtseva O. M., Akimov L. A., Korokhin V. V. 2002, Icarus, 157, No2, 419-425.
\reference Hall J.S. and Riley L.A. 1974, T. Gehrels. Ed., University of Arizona Press, Tucson, Arizona, 593-598.p.1;
\reference Beebe R.F., Suggs R.M., Little T. 1986, Icarus, 66, 359-365.
\reference Caldwell J, Cess R., Carlson B. 1979, \apj, 234, 155-158
\reference West R. A. 1988, Icarus, 75, 381-398
\reference Xanthakis J. et al. 1980, $\Pi\rho\alpha\kappa\tau\iota\kappa\alpha$ $\tau\eta\sigma$ $\alpha\kappa\alpha\delta\eta\mu\tau\alpha\sigma$
$\alpha\vartheta\eta\nu\omega\nu$, 55, 362-371
\end {references}
\end{document}